\documentclass[12pt, a4paper, oneside]{article}
\usepackage{pythonhighlight}
\usepackage{xcolor} 
\usepackage{algorithm}
\usepackage{algpseudocode}
\usepackage{listings}
\usepackage{authblk} 
\usepackage{hyperref} 
\usepackage{quantikz}
\usepackage{amssymb}
\usepackage[utf8]{inputenc}
\DeclareUnicodeCharacter{030C}{\v{}}
\usepackage{mathtools}
\usepackage{gensymb}
\usepackage{textcomp}
\usepackage{multicol}
\usepackage{blindtext}
\usepackage{float}
\usepackage[bb=px]{mathalfa} 
\usepackage[dvipsnames]{xcolor}
\usepackage{hyperref}
\usepackage{enumitem}
\usepackage{multirow}   
\usepackage{array} 
\usepackage{makecell}
\hypersetup{ linktocpage,
     colorlinks=true, 
     linkcolor= teal,  
     citecolor = blue,       
     urlcolor=black, 
     }     
\usepackage{amsmath}
\usepackage{amsthm}
\usepackage{svg}
\usepackage{graphicx}

\usepackage{comment}
\usepackage[english]{babel}
\usepackage{subcaption}
\usepackage[utf8]{inputenc}
\usepackage[a4paper,
            left=0.75in,
            right=0.75in,
            top=0.75in,
            bottom=0.75in,
            footskip=.5in]{geometry}
\usepackage{csquotes}
\usepackage[
    backend=biber,
    style=numeric-comp,
    sorting=none,
    maxbibnames=99,      
    doi=false,
    url=false,
    isbn=false,
    eprint=false
]{biblatex}

\newtheorem{theorem}{Theorem}

\title{Single-Round Deterministic \\ Quantum Anonymous Veto Using Bell States} %Efficient Single-Round QAV Protocol with Bipartite Entanglement

\author[1]{Ravi Sangwan} %\thanks{\href{mailto:phy.ravi9@gmail.com}{\texttt{phy.ravi9@gmail.com}}}}
\author[1]{Harishankar Mishra} %\thanks{\href{mailto:harishankarm@cdac.in}{\texttt{harishankarm@cdac.in}}}}
\author[1]{Henry Sukumar } %\thanks{\href{mailto:henrys@cdac.in}{\texttt{henrys@cdac.in}}}}
\author[1]{Gudapati Naresh Raghava\thanks{%Correspondence: 
\href{mailto:nareshraghava@cdac.in}{\texttt{nareshraghava@cdac.in}}}}

\affil[1]{Quantum Technology Group, Centre for Development of Advanced Computing (C-DAC), Bangalore, India.}

\date{}
\begin{document}
\maketitle 
\renewcommand{\abstractname}{}
\begin{abstract}
Quantum anonymous veto protocols provide a secure framework for group decision-making, while preserving participant privacy. Existing approaches often rely on multipartite entanglement or require multiple communication rounds, limiting their practical applicability. In this work, we propose a deterministic quantum anonymous veto protocol based solely on Bell states that operates in a single communication round. By employing a binary phase encoding strategy across multiple entangled pairs, the protocol achieves efficient aggregation of veto information with logarithmic scaling in entanglement resources. The protocol preserves all key security properties, including anonymity, privacy, binding, correctness, commutativity, verifiability, and robustness, which are formally verified through a detailed security and efficiency analysis. Furthermore, we outline a feasible photonic implementation using standard optical components. The proposed approach provides a scalable and experimentally viable framework for anonymous decision-making in distributed quantum systems.

\end{abstract}

%\clearpage
%\tableofcontents
%\thispagestyle{empty}
%\clearpage
\pagenumbering{arabic} 
\section{Introduction}
\label{sec:introduction}

Collective decision-making is a core aspect of modern distributed communication systems, collaborative networks, and multi-party governance frameworks that is secured. In many such settings, a decision is accepted only if no member opposes it. This gives rise to the issue of an anonymous veto, where the group needs to figure out whether or not one member has vetoed a proposal and at the same time make sure that the identity of the vetoing member is unknown at all. Designing a protocol that ensures anonymity, privacy,  correctness, binding, verifiability, and resistance to malicious behaviour is a hard task, especially when the participants do not trust one another completely. Classical anonymous veto protocols rely on computational assumptions and trusted authorities, and therefore cannot provide unconditional security. Moreover, their multi-round structure increases communication overhead and the risk of information leakage. These limitations motivate the use of quantum communication techniques.

Quantum information science provides an fundamentally new framework of secure communication through the use of non-classical properties of quantum states and communication protocols, namely superposition, entanglement, and the no-cloning property, respectively~\cite{Nielsen_Chuang_2010,BENNETT20147,ekert1991,gisin2002,scarani2009,xu2020}. These principles enable cryptographic primitives and voting processes whose security is unconditionally guaranteed with quantum methodology~\cite{hillery2006,vaccaro2007,thapliyal2017,wang2016,jiang2012,bonanome2011,xue2017,sun2019,jiang2020,liu2021a,wang2020,wang2021a,du2021,shi2021,sekga2021,liu2021b,li2020,li2021,zhang2020,joy2020,mishra2025deterministic}. The Quantum Anonymous Veto (QAV) protocols are a more specific subgroup of quantum voting schemes with the advantage that allow participants to determine whether at least one veto has been cast, without revealing the identity.

Rahman and Kar  introduced the first formal QAV protocol, which is commonly known as the RKQAV protocol~\cite{rahman2015}. This protocol uses an $n$-qubit Greenberger-Horne-Zeilinger (GHZ) state which is shared by the $n$ participants. The qubits are measured to see whether one of the vetoes has been cast and the participants encode their choice by utilizing a local unitary operation. The protocol is implemented in several rounds, with different GHZ states being generated and distributed in each round. The amount of rounds necessary is restricted by $1 + \log_2 n$, which is an indication of recursive querying strategy. Although the RKQAV protocol provides privacy and correctness, it lacks binding and robustness and suffers from severe scalability limitations. The scaling of large GHZ states and their preparation and distribution is more complicated with increasing the number of participants, and the sensitivity of multipartite entanglement to the protocol difficulties the scaling. Moreover, the necessity to perform several rounds of communication complicates the consumption of the resources and the complexity of the operations.

To overcome the multi-round limitation of RKQAV, Wang \emph{et al.} proposed a probabilistic single-round QAV protocol (WQAV) based on multiple copies of GHZ states~\cite{wang2021}. In this protocol, the various members are given one qubit of a number of copies of an n-qubit GHZ state and encode their choice with local operations. Once the encoding phase is complete, the qubits are sent back to the voting authority, which then makes collective measurements to determine whether or not a veto was cast. Even though WQAV provides privacy, binding, and verifiability and minimizes communication rounds, it lacks robustness and compromises deterministic correctness. Many GHZ state copies are needed to obtain some high confidence of the result, which grows the quantum communication overhead. Further, the dependence on multipartite entanglement still is a major experimental problem and it inhibits the feasibility of the protocol on a larger scale.

Subsequently, Mishra et al.~\cite{mishra2022} introduced a set of seven QAV protocols. Among these, the iterative Bell-state-based scheme (referred to as QAV-6) and the deterministic multi-qubit entangled-state scheme (QAV-7) were identified by the authors as the most efficient. Specifically, the Bell-state based protocol is iterative and substitutes multipartite GHZ states by Bell pairs which are much easier to prepare and transmit in realistic quantum systems. In this protocol a Bell state is prepared by the voting authority who keeps back one qubit (the home qubit), and transmits the travel qubit to the other participants sequentially (there being a total of n participants). A decision is picked by each participant as a phase operation to the travel qubit. Once the qubit has been through all the participants, the qubit is sent back to the voting authority who then measures the qubit in the Bell-basis to decide whether there is a veto or not. The protocol is iterative, and it take up to $1 + log_2n$ rounds of communication to arrive at a definitive answer. The transmission of quantum states through many rounds raises the likelihood of decoherence and  noise in channels. Furthermore, the protocol does not strictly satisfy the binding property, as participants may alter their decisions in subsequent rounds.

In contrast, in the deterministic multi-qubit entangled-state protocol the voting authority prepares a multi-qubit entangled state (a cluster state or GHZ-type state) with $m$ qubits , where \(m \leq n\), with \(n\) representing the number of participants, and then sends subsets of qubits to the participants. Each participant performs a unitary operation based on his or her decision and the final state is measured to establish whether or not a veto was cast. The protocol has a deterministic single-round veto detection and anonymity. Nevertheless, there are experimentally challenging processes to generate and distribute large multi-qubit entangled states. Moreover, the complexity of the design of participant-specific unitary operations increases with the number of participants, and it can be challenging to implement a large-scale operation.

In general, the existing QAV protocols can be classified into two dimensions: (i) the entanglement structure used (multipartite and bipartite) and (ii) operational determinism (iterative and single-round). Multipartite-entanglement-based protocols have severe scalability limitations as well as resource acquisition limitations, while iterative protocols are less resource-efficient and do not satisfy the binding property. This gives rise to a fundamental trade-off: designing a single-round deterministic QAV protocol using bipartite entanglement.

To address these limitations, this work introduces  a modified deterministic QAV protocol, inspired by Bell-state-based iterative protocol (QAV-6), which completes the veto detection in
a single round using only bipartite entanglement in the form of Bell states. To the best of our knowledge, it is the first QAV protocol that simultaneously achieves deterministic veto detection, single-round communication, and implementation using only bipartite entanglement (Bell states), with logarithmic resource scaling in the number of participants.

All the security properties, including anonymity, privacy, correctness, verifiability, robustness, commutativity, and binding that are analyzed in this work, are upheld by the protocol. Furthermore, we provide a photonic implementation framework based on polarization-path encoding and discrete-time quantum walks (DTQWs), outlining how the protocol can, in principle, be realized using current quantum optical hardware.

Along with the theoretical contribution, the proposed protocol has significant practical implications. It supports the secure distributed decision making where the privacy of the participants is important such as aanonymous committee voting in decentralized organizations, privacy-preserving
consensus mechanisms, and quantum-secure governance system By combining scalability with strong security guarantees and experimental feasibility, the proposed scheme provides a viable pathway toward real-world deployment of quantum anonymous voting protocols.

The rest of this paper is structured as follows. Section~\ref{sec:proposed_protocol} describes the proposed deterministic Bell-state-based QAV protocol in detail. Section~\ref{sec:security_efficiency} presents a comprehensive security and efficiency analysis. Section~\ref{sec:implementation_using_quantum_walk} outlines a photonic implementation framework using discrete-time quantum walks. Finally, Section~\ref{sec:conclusion} summarizes the main contributions of the work and outlines possible directions for future research.

\section{Deterministic QAV Protocol Using Bell States}
\label{sec:proposed_protocol}

This section introduces the proposed deterministic Quantum Anonymous Veto (QAV) protocol, which enables anonymous detection of vetoes using Bell states within a single iteration. 

\subsection{Model and Assumptions}
We consider a sequential quantum anonymous veto protocol involving \(n\) participants denoted by \(V_i\) for \(i \in \{0,1,\ldots,n-1\}\), each possessing the ability to either approve or veto a proposal. Let \(k\) denote the number of participants who choose to exercise their veto authority, where \(0 \le k \le n\).  A voting authority (VA) executes the protocol, which includes the preparation of maximally entangled quantum states and distributes qubits to the participants and finally performs the bell state measurements to determine the presence of any veto.  Bell states of the form  \[ |\Phi^+\rangle = \frac{1}{\sqrt{2}}(|00\rangle + |11\rangle) \]  are employed by the protocol. For each Bell pair, the VA retains one qubit referred to as the  \emph{home qubit} \(h_a\), while the second qubit, called the \emph{travel qubit} \(t_a\), is transmitted through the participants.  To reliably identify the existence of vetoes, the protocol requires the use of  \begin{equation*} h = \lfloor \log_2 n \rfloor + 1 \label{Eq:h bell states} \end{equation*}  Bell states. Every Bell pair is a phase probe, which is associated with the position of a bit in the binary representation of the veto count $k$.  The travel qubits are subjected to controlled phase operations by which the participants encode their voting decision.

We assume a semi-honest VA (honest-but-curious) that follows the prescribed protocol strictly but could make an effort of extracting ancillary information out of the accessible quantum states and measurement results. The participants are assumed to be computationally unbounded but constrained by the laws of quantum mechanics. Based on these assumptions, the deterministic QAV protocol operates as described in next subsection. 

 \subsection{Protocol} \label{sec: protocol} The key to the protocol is the encoding of the veto number in the relative phase of each Bell pair with controlled phase rotations. Each Bell pair acts as a probe of a certain bit position on the binary representation of $k$. This coded information is extracted by the voting authority by Bell-basis measurement, thus making it possible to deterministically detect vetoes on the basis of using only $h$ Bell states in a single round of operation.

The steps of the protocol are as follows:

\begin{enumerate}
    \item {Initialisation:}  
    The VA prepares \( h \) maximally entangled Bell states of the form
    \[
    |\Phi^+\rangle = \frac{1}{\sqrt{2}} (|00\rangle + |11\rangle),
    \]
    retaining one qubit (the \emph{home qubit} \( h_a \)) from each pair and distributing the corresponding \emph{travel qubit} \( t_a \) to the first participant \( V_0 \), for each index \( a \in \{1, 2, \ldots, h\} \).

    \item {Sequential Vetoing:}  
    Each travel qubit \( t_a \) sequentially passes through all participants \( V_i \). Each participant performs the following local operation:
    \begin{itemize}
        \item If the participant chooses to \emph{veto}, they apply the phase gate:
        \[
        P^{(a)} = 
        \begin{pmatrix} 
        1 & 0 \\ 
        0 & e^{i \pi / 2^{a-1}} 
        \end{pmatrix}, 
        \quad a \in \{1, 2, \ldots, h\}.
        \]
        \item Otherwise, they apply the identity operation.
    \end{itemize}
    Since these phase gates commute, the order of application does not affect the final state. After all participants have acted on each qubit, the travel qubits are returned to the VA.

    \item {Final Measurement:}  
    The VA performs a Bell basis measurement on each home–travel qubit pair. Initially, the joint state is \( |\Phi^+\rangle \). If one or more participants apply a phase gate to \( t_a \), the resulting state becomes
    \[
    |\Phi^{(a)}_k\rangle = 
    \frac{1}{\sqrt{2}} 
    (|00\rangle + e^{i k \pi / 2^{a-1}} |11\rangle),
    \]
    where \( k \) denotes the total number of vetoes. This state corresponds to a valid Bell state if and only if
    \[
    \frac{k \pi}{2^{a-1}} \equiv 0 \pmod{2\pi} 
    \quad \text{or} \quad 
    \pi \pmod{2\pi}.
    \]
    Consequently, the measurement outcome is:
    \begin{itemize}
        \item \( |\Phi^+\rangle \) if \( \frac{k \pi}{2^{a-1}} \equiv 0 \pmod{2\pi} \),
        \item \( |\Phi^-\rangle \) if \( \frac{k \pi}{2^{a-1}} \equiv \pi \pmod{2\pi} \).
    \end{itemize}

    Specifically:
    \begin{itemize}
        \item The 1st Bell pair (\( a = 1 \)) yields \( |\Phi^-\rangle \) iff \( k \) is odd.
        \item The 2nd Bell pair (\( a = 2 \)) yields \( |\Phi^-\rangle \) iff \( k \equiv 2 \pmod{4} \).
        \item The 3rd Bell pair (\( a = 3 \)) yields \( |\Phi^-\rangle \) iff \( k \equiv 4 \pmod{8} \).
        \item In general, the \( a^{th} \) Bell pair yields \( |\Phi^-\rangle \) if and only if the \( a^{th} \) least significant bit of \( k \) (in binary) is 1 and all less significant bits are 0.
    \end{itemize}
    
    \item {Veto Detection:}  
    The VA infers the presence of a veto from the Bell measurement outcomes:
    \begin{itemize}
        \item If all \( h \) outcomes are \( |\Phi^+\rangle \), then \( k = 0 \), indicating no veto.
        \item If any outcome is \( |\Phi^-\rangle \), then \( k \geq 1 \), indicating that at least one veto was cast.
    \end{itemize}
    This detection process preserves anonymity by revealing only the presence of vetoes, not the identity or number of vetoing participants.
\end{enumerate}

The entire procedure is summarized in Algorithm~\ref{alg:qav}.

\begin{algorithm}[H]
\caption{Deterministic QAV Protocol Using Bell States}
\label{alg:qav}
\begin{algorithmic}[1]
\Require $n$ participants $V_0, V_1, \ldots, V_{n-1}$
\Ensure Detect presence of veto ($k \geq 1$) without revealing participant identities

\State Compute $h \gets \lfloor \log_2 n \rfloor + 1$
\For{$a = 1$ to $h$}
    \State VA prepares Bell state $|\Phi^+\rangle = \frac{1}{\sqrt{2}}(|00\rangle + |11\rangle)$
    \State VA retains home qubit $h_a$ and sends travel qubit $t_a$ to $V_0$
\EndFor

\For{$i = 0$ to $n - 1$}
    \For{$a = 1$ to $h$}
        \If{$V_i$ chooses to veto}
            \State Apply $P^{(a)} = 
            \begin{pmatrix} 
            1 & 0 \\ 
            0 & e^{i\pi/2^{a-1}} 
            \end{pmatrix}$ to $t_a$
        \Else
            \State Apply identity operation to $t_a$
        \EndIf
    \EndFor
    \State Forward all $t_a$ to $V_{i+1}$ \Comment{or return to VA if $i = n-1$}
\EndFor

\For{$a = 1$ to $h$}
    \State VA performs Bell basis measurement on $(h_a, t_a)$
    \State Record outcome as $|\Phi^+\rangle$ or $|\Phi^-\rangle$
\EndFor

\If{All outcomes are $|\Phi^+\rangle$}
    \State \Return No veto ($k = 0$)
\Else
    \State \Return Veto detected ($k \geq 1$)
\EndIf
\end{algorithmic}
\end{algorithm}

\begin{figure}[H]
    \centering
    \includegraphics[width=0.8\linewidth]{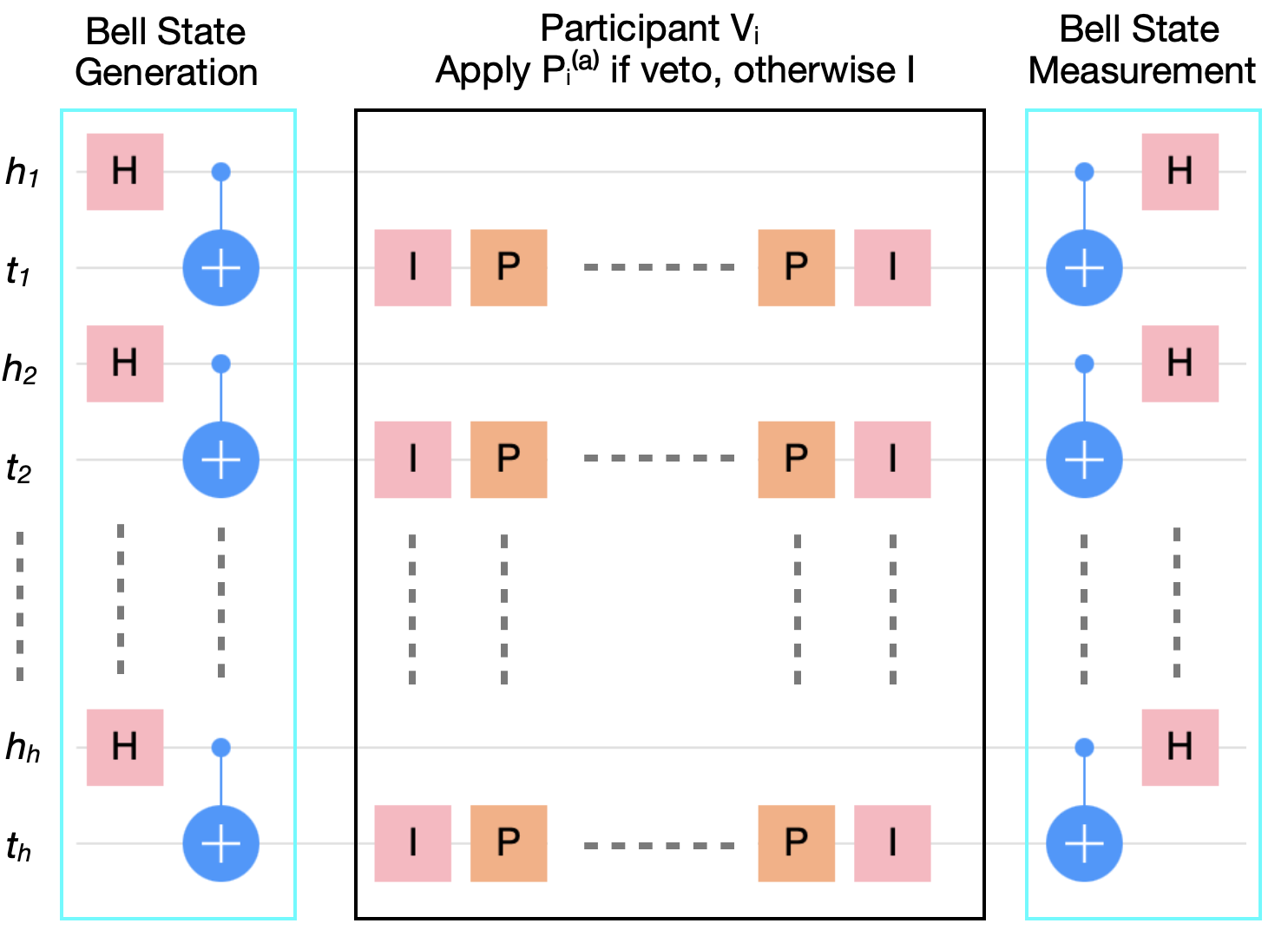}
\caption{Generic circuit for the deterministic QAV protocol with $h$ Bell pairs and $n$ participants. The VA prepares $h$ Bell states~\eqref{Eq:h bell states} $(h_a,t_a)$, retains home qubits $h_a$, and sends travel qubits $t_a$ sequentially through participants $V_i$, each applying veto-dependent phase operations $P_i^{(a)}$. The travel qubits are returned to the VA for Bell-basis measurement.}
\label{fig:qav_generic_circuit}
\end{figure}

\begin{figure}[H]
    \centering
    \includegraphics[width=0.8\linewidth]{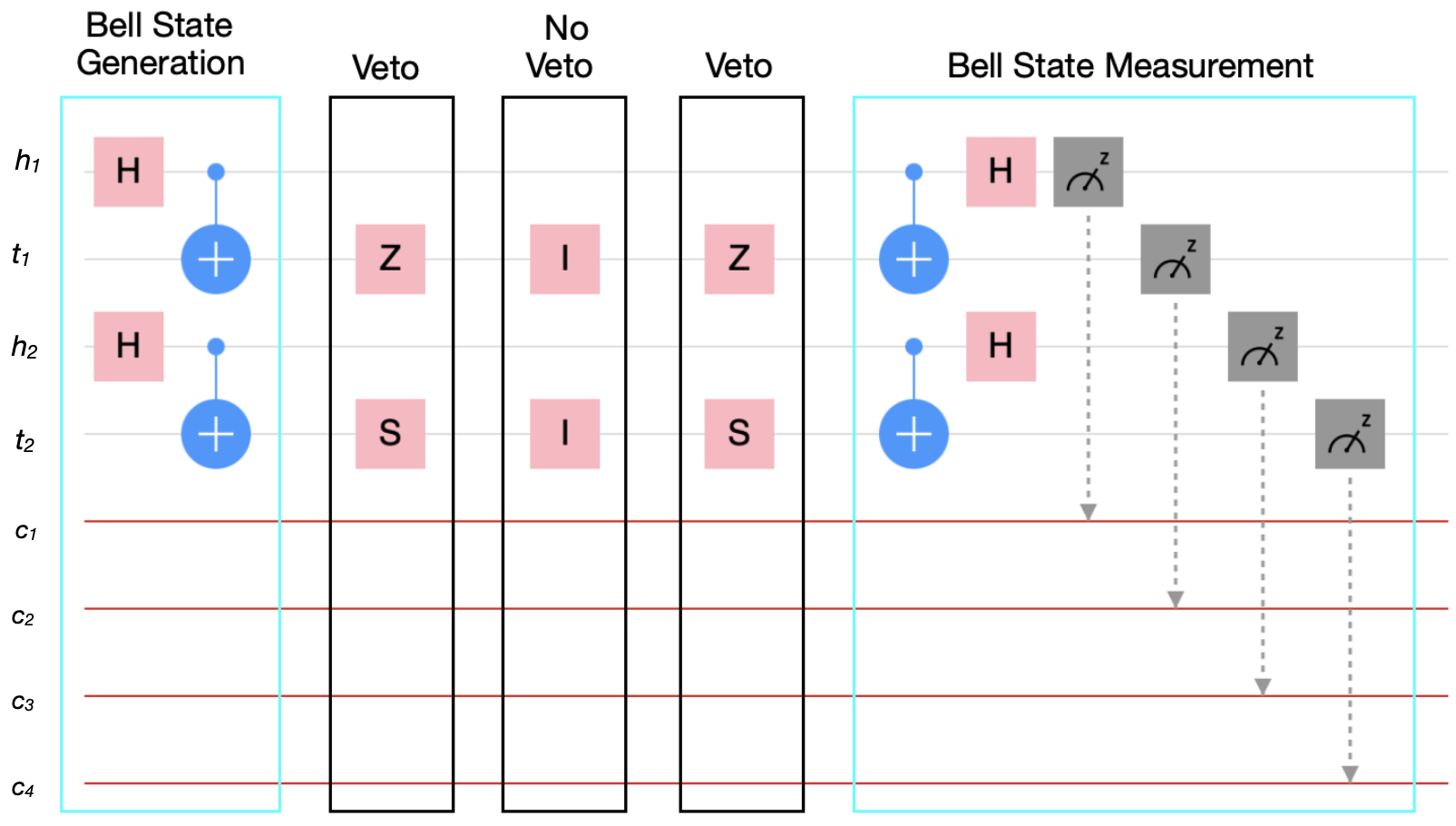}
    \caption{Quantum circuit demonstrating of the proposed algorithm for $3$ participant where the first and the third participant exercises their veto authority.}
    \label{fig : qcirc}
\end{figure}

The quantum circuit representation of the proposed protocol as described in Algorithm~\ref{alg:qav} is shown in Figure~\ref{fig:qav_generic_circuit}. In addition, Figure~\ref{fig : qcirc} illustrates a specific instance of the protocol for three participants, where the first and the third participants exercise their veto authority. The VA generates $2$ Bell states~\eqref{Eq:h bell states} $|\Phi^+\rangle$ and then passes one of the bell pair qubits from each of the bell states ($t_a$) to the first participant. Then each participant applies the unitary operator based on their decision of veto and sends the $t_a$ to the next participant and finally to the VA. VA then performs the bell state measurement and based on the measurement results they concludes whether someone has exercise their veto authority or not.

The protocol guarantees \emph{correctness}, since a veto is detected if and only if \( k > 0 \); \emph{anonymity}, as the VA learns only the existence but not the identity of vetoing participants; and \emph{determinism}, as the process completes in a single round of entanglement distribution, vetoing, and measurement. These properties are analyzed in detail in the next section, where we evaluate the proposed scheme in terms of its security guarantees and qubit efficiency.

 %The security analysis encompasses both intrinsic guarantees arising from the protocol design and supplementary mechanisms based on quantum digital signatures and decoy states that enhance authentication and channel integrity.
%Moreover, the use of quantum digital signature and decoy states in order to prepare subsequent subroutines can be adopted to guarantee authenticate participant involvement and secure quantum communication channels.

\section{Security and Efficiency Analysis}
\label{sec:security_efficiency}

In this section, we present a comprehensive evaluation of the proposed deterministic QAV protocol in terms of its security guarantees and qubit efficiency. 

\subsection{Security Analysis} \label{sec:security_gaurantees}
The proposed deterministic Bell-state based QAV protocol satisfies the essential security requirements expected in quantum vetoing frameworks, including eligibility, privacy, binding, correctness, verifiability, and robustness. These properties follow from the structure of the protocol, the use of authenticated participants, and the deterministic encoding of veto information in the relative phases of Bell states. 
For the threat model, we assume that VA follows the prescribed protocol but may attempt to infer additional information from the accessible quantum states and the final measurement outcomes. Participants are allowed to perform arbitrary local quantum operations on the travelling qubit according to the protocol. External adversaries may attempt intercept, resend or disturbance attacks during transmission.

\begin{enumerate}
    \item {Eligibility:} 
     In order to secure authenticated veto, the VA can employ a quantum digital signature scheme based on BB84 states ~\cite{Quantum_Sign}. Each participant \(V_i\) registers with the VA by sending an extended sequence of BB84 quantum states \((|0\rangle, |1\rangle, |+\rangle, |-\rangle)\). The VA samples the qubits randomly in the computational or Hadamard basis, creating an eliminated signature that is specific to an individual participant. For instance, when the VA measures a qubit and finds the result of measuring to be \(|+\rangle\), it can remove the possibility of state \(|-\rangle\), as the two are orthogonal. The elimination results obtained from each qubit are combined to form a partial quantum signature. Under the vetoing phase, every participant reveals the initially prepared sequence and the VA checks it against the eliminated signature. In case the rate of mismatch is less than that which has been set, the participant is verified. This is done to make sure that only bonafide and registered participants are able to participate in the protocol. Thus, the protocol guarantees eligibility.
    \item {Privacy and Anonymity:}  
    The protocol guarantees privacy against a semi-honest VA, the malicious participants, and external eavesdroppers. Initially, each Bell pair prepared by the VA is in the state
    \[
    |\Phi^+\rangle = \frac{1}{\sqrt{2}} (|00\rangle + |11\rangle),
    \]
    with density matrix
    \[
    \rho_{ht}^{(0)} = |\Phi^+\rangle \langle \Phi^+|.
    \]
    If a total of $k$ participants apply the phase operation on the travel qubit, the joint state becomes
    \[
    |\Phi_k^{(a)}\rangle =
    \frac{1}{\sqrt{2}}
    \left(|00\rangle + e^{i k \pi / 2^{a-1}} |11\rangle \right),
    \]
    with density matrix
    \[
    \rho_{ht}^{(k)} = |\Phi_k^{(a)}\rangle \langle \Phi_k^{(a)}|.
    \]
    
    The reduced density matrix accessible to any participant or external adversary is obtained by tracing out the home qubit:
    \[
    \rho_t^{(k)} = \mathrm{Tr}_h \left( \rho_{ht}^{(k)} \right)
    = \frac{1}{2} I,
    \]
    which is maximally mixed and independent of $k$ and of the identity of the vetoing participants. 
    No local measurement can reveal individual veto decisions, and the VA’s Bell-basis measurement reveals only the global veto condition ensuring privacy. 
    The anonymity property of the protocol is formally shown in Theorem~\ref{thm:anonymity}.  \begin{theorem} \label{thm:anonymity} Let $\mathit{v} = (v_1,v_2,\dots,v_n)$ and  $\mathit{v'} = (v'_1,v'_2,\dots,v'_n)$ be two voting configurations such that \[ \sum_{i=1}^{n} v_i = \sum_{i=1}^{n} v'_i = k .\] Then the final quantum states produced by the protocol are identical, \[ \rho_{\mathit{v}} = \rho_{\mathit{v'}} . \] \end{theorem} \begin{proof} Each participant applies either the identity operation or the phase gate $P^{(a)}$ on the corresponding travelling qubit $t_a$. Since all phase operations commute, the  overall unitary applied by all participants become \[ U = \prod_{i=1}^{n} (P^{(a)})^{v_i} = (P^{(a)})^k , \] where $k = \sum_{i=1}^{n} v_i$ is the total number of vetoes. Let the initial Bell state prepared by the voting authority be \[ |\Phi^+\rangle = \frac{1}{\sqrt{2}} (|00\rangle + |11\rangle). \] After all participants apply their operations, the final encoded state becomes \[ |\Phi_k^{(a)}\rangle = (I \otimes (P^{(a)})^k)|\Phi^+\rangle . \] Thus, the final state depends only on the total number of vetoes $k$ and not on  which participant applied the phase operation. Therefore, \[ |\Phi_{\mathit{v}}\rangle = |\Phi_{\mathit{v'}}\rangle , \] which implies equality of the corresponding density matrices. Hence, no measurement performed by the voting authority or any adversary can distinguish  which participant cast the veto. \end{proof}
    
    Privacy against participant-side coherent attacks can also be shown formally. Suppose a dishonest participant attaches an ancillary system $A$ initialized in state $|0\rangle_A$ and applies an arbitrary joint unitary operation $U_{tA}$ on the travel qubit $t$ and the ancilla. The combined state becomes
    \[
    \rho_{htA}^{(k)} = U_{tA} \left( \rho_{ht}^{(k)} \otimes |0\rangle\langle 0|_A \right) U_{tA}^\dagger.
    \]
    The information accessible to the adversary is described by the reduced state
    \[
    \rho_{tA}^{(k)} = \mathrm{Tr}_h \left( \rho_{htA}^{(k)} \right).
    \]
    Since the reduced state of the travel qubit alone satisfies $\rho_t^{(k)} = \frac{1}{2} I$ for all values of $k$, any joint operation on $t$ produces a state $\rho_{tA}^{(k)}$ that is independent of the identity of the vetoing participants. Moreover, because the reduced state remains maximally mixed regardless of previous phase operations, no participant can extract information about earlier votes through local measurements. Ancillary entanglement attacks do not provide useful information, since the marginal state is independent of the applied phase operations.
    
    To protect against external eavesdropping during transmission, the VA randomly adds decoy states, which can either be decoy states of the BB84 basis, or the maximally entangled Bell states to each sequence of the transmitted qubits~\cite{decoy_qubits}. Once transmitted, the VA discloses bases of these decoy states, and each receiving participant may then measure each of them. When the measured quantum bit error rate goes above some predefined threshold, it is inferred that the channel has been compromised by eavesdropping or affected by noise. Decoy state method allows early identification of attacks and as such, veto transmission can only be done with use of safe quantum channels. This in conjunction with the quantum digital signature based authentication will go along way in ensuring the security of the system against most of the external and internal attackers.

    Each participant's choice remains perfectly concealed. The only allowed local operations, identity or fixed phase rotation do not reveal individual intentions. The VA's final Bell-basis measurement reveals solely the global veto condition (\(k \ge 1\)) without disclosing which participant applied the phase poeration and the protocol guarantees unconditional privacy of participant choices.
    \item {Binding:}  
    Once a participant has operated on the travel qubits, their choice is irreversible. Since the qubits propagate unidirectional and each local operation is unitary, no participant can later modify their encoded decision.
    \item {Correctness:}  
    Each participant $P_i$ encodes their veto decision $v_i \in \{0, 1\}$ by applying the phase operation $P^{(a)}$ on the corresponding \emph{travel qubit} \( t_a \). Since all phase operations commute, the overall operation applied by all participants becomes \begin{equation*} U = \prod_{i=1}^{n} \left(P^{(a)}\right)^{v_i} = \left(P^{(a)}\right)^k, \end{equation*} where \begin{equation*} k = \sum_{i=1}^{n} v_i  \end{equation*}
    The initial Bell state prepared by the VA is \begin{equation*} |\Phi^+\rangle = \frac{1}{\sqrt{2}} (|00\rangle + |11\rangle). \end{equation*} After all participants apply their phase operations, the final state becomes \begin{equation*} |\Phi_k^{(a)}\rangle = (I \otimes (P^{(a)})^k)|\Phi^+\rangle. \end{equation*}
    The Bell-basis measurement performed by the VA distinguishes whether $k = 0$ (no veto) or $k \ge 1$ (presence of veto), ensuring deterministic correctness of the protocol.
    \item {Information Leakage:}  The analysis of correctness indicates that the Bell-basis measurements carried out by the VA determine only whether or not the aggregate veto condition satisfies or not  (\(k = 0\) or \(k \ge 1\)), where \(k\) denotes the total number of veto  operations applied by the participants. More importantly, the measurement results are solely based on the collective phase accumulated from all participant operations and do not reveal who the vetoing participant(s) is as well as the exact amount of vetoes. Thus, the protocol discloses the desired overall decision value, but maintains the anonymity of the individual participant decisions. This property will ensure that the protocol has a controlled leakage of the information necessary to the decision-making process.
    \item {Verifiability:}  
    After completing the protocol, the semi honest VA publicly announces the Bell-basis measurement outcomes for all Bell pairs. Since these outcomes depend only on the aggregate phase accumulated from all participants’ operations, each participant can independently verify the correctness of the announced result using the public information and their own local operations. This verification process does not reveal the identity of vetoing participants, as the measurement outcomes encode only the global veto condition and not the individual contributions. Therefore, the protocol ensures verifiability without compromising anonymity.
    \item {Commutativity:} The proposed protocol possesses the property of order independence meaning the final output does not depend on the order in which the participants execute their operations. The decision of each of the participants in encoded into the travelling qubit of the $a^{\text{th}}$ Bell pair using the phase operation \[P^{(a)} = \begin{pmatrix} 1 & 0 \\ 0 & e^{i\pi/2^{a-1}} \end{pmatrix}.\]. Since this is a diagonal operation, all such phase operation will commute with another. For any two participants $V_i$ and $V_j$, \[P^{(a)}_i P^{(a)}_j = P^{(a)}_j P^{(a)}_i.\] and the composite operation on the travelling qubit is only dependent on the total number of veto operations, and not on the order in which they are executed. When the veto operation is used on the $a^{\text{th}}$ travelling qubit by the $k$ number of participants, the total operation then becomes. \[ (P^{(a)})^{k} = \begin{pmatrix}1 & 0 \\0 & e^{ik\pi/2^{a-1}} \end{pmatrix}.\] Applying this to the initial Bell state
\[
|\Phi^+\rangle = \frac{1}{\sqrt{2}}\left(|00\rangle + |11\rangle\right)
\]
produces the encoded state
\[
|\Phi_k^{(a)}\rangle = \frac{1}{\sqrt{2}} \left( |00\rangle + e^{ik\pi/2^{a-1}} |11\rangle \right).
\]
In this way the final state is just a function of the number of vetoes $k$ as well as independent of the sequence in which participants perform their operations. Consequently, the protocol is order-independent and fair.

\item {Robustness:}
Application of quantum communication protocols into practice is inevitably influenced by decoherence that is induced by the interaction with the surrounding world. Even in absence of malicious activity, environmental noise that operates on transmitted qubits can distort information and  cause incorrect results. Thus, it is important to consider the performance of the protocol in the conditions of the noise.

The travel qubits are sent in turn among participants in the proposed deterministic QAV protocol. We thus model the noise as only taking action on the travel qubits in the transmission between the participants. The operator-sum representation, used to describe the effect of noise on a quantum state is
\[
\rho_f = \sum_i E_i \rho_i E_i^\dagger,
\]

where $\{E_i\}$ are Kraus operators~\cite{Nielsen_Chuang_2010} satisfying $\sum_i E_i^\dagger E_i = I$.

After the encoding operations of all participants, the ideal state of the 
$a^{\text{th}}$ Bell pair is

\[
|\Phi^{(a)}_k\rangle =
\frac{1}{\sqrt{2}}
\left(|00\rangle + e^{i k \pi / 2^{a-1}} |11\rangle \right).
\]

Since the travel qubit passes sequentially through all $n$ participants, it 
experiences noise $n$ times. Let $\mathcal{E}$ denote the noise channel acting 
on the travel qubit. The final noisy state becomes

\[
\rho^{(a)}_k =
(I \otimes \mathcal{E}^{\,n})
\left(
|\Phi^{(a)}_k\rangle \langle \Phi^{(a)}_k|
\right).
\]

To quantify the effect of noise, we compute the fidelity between the noisy state 
and the ideal encoded state,

\[
F^{(a)} =
\langle \Phi^{(a)}_k |
\rho^{(a)}_k
| \Phi^{(a)}_k \rangle,
\]

which gives the probability of correctly identifying the Bell-basis outcome.

For amplitude damping noise with parameter $\lambda$, the Kraus operators~\cite{Nielsen_Chuang_2010} are
\[
E_0 =
\begin{pmatrix}
1 & 0 \\
0 & \sqrt{1-\lambda}
\end{pmatrix},
\quad
E_1 =
\begin{pmatrix}
0 & \sqrt{\lambda} \\
0 & 0
\end{pmatrix}.
\]

Repeated application over $n$ transmissions reduces the coherence between 
$|00\rangle$ and $|11\rangle$ as $(\sqrt{1-\lambda})^n$. Substituting into the 
fidelity expression yields

\[
F^{(a)}_{\lambda}
=
\frac{1}{2}
\left(
1 + (1-\lambda)^{n/2}
\right).
\]

For phase damping noise with parameter $\gamma$, the Kraus operators are

\[
E_0 =
\begin{pmatrix}
1 & 0 \\
0 & \sqrt{1-\gamma}
\end{pmatrix},
\quad
E_1 =
\begin{pmatrix}
0 & 0 \\
0 & \sqrt{\gamma}
\end{pmatrix},
\]
Phase damping suppresses only the off-diagonal terms of the density matrix. 
After $n$ transmissions, the coherence term is reduced by $(1-\gamma)^n$, giving

\[
F^{(a)}_{\gamma}
=
\frac{1}{2}
\left(
1 + (1-\gamma)^n
\right).
\]
Since the veto information is only encoded in the relative phase between the basis states, phase damping has a more impact than amplitude damping.
 
The overall robustness of the protocol is determined by the average fidelity 
over all $h$ Bell pairs,

\[
F_{\mathrm{avg}}
=
\frac{1}{h}
\sum_{a=1}^{h}
F^{(a)}.
\]
 
For small values  of the noise parameters fidelity is near to unity, meaning the protocol can be used with a high degree of reliability to identify when a veto occurs in the presence of moderate environmental disturbances. With a higher noise strength, the loss of coherence decreases the distinguishability of the Bell states, which ultimately limits the performance of the protocol. These findings suggest that the suggested deterministic QAV protocol can be used with stable performance in the case of moderate amplitude and phase damping. \end{enumerate}

Tables~\ref{tab:1} and~\ref{tab:2} provide a comparative summary of the security properties and resource efficiency of QAV-6, QAV-7, and the proposed protocol.

\subsection{Qubit Efficiency Analysis}
The performance of quantum communications protocols can be measured by qubit efficiency~\cite{qubit_effi}, defined as
\begin{equation*}
    \eta = \frac{c}{q + b},
    \label{eq:efficiency_def}
\end{equation*}
where \(c\) denotes the number of classical bits communicated as the final output, \(q\) is the total number of qubits transmitted, and \(b\) is the number of classical bits exchanged for secure transmission (excluding those used for eavesdropping detection). For QAV protocols, \(c = 1\), since the final outcome is a single classical bit indicating whether a veto has been cast.

The VA prepares
\begin{equation*}
    h = \lfloor \log_2 n \rfloor + 1
\end{equation*}
Bell pairs for \(n\) participants. For each Bell pair, the home qubit remains with the VA, while the travel qubit sequentially passes through all \(n\) participants before returning to the VA for Bell-basis measurement.

Each hop of the travel qubit between two participants or between a participant and the VA is modelled as an independent insecure quantum channel. Accordingly, for each hop, \(\delta_1\) decoy qubits are appended for eavesdropping detection. The qubit cost per Bell pair is therefore
\begin{equation*}
    q_{\text{per Bell pair}} = (n + 1)(1 + \delta_1) + 2,
\end{equation*}
where \((n + 1)\) accounts for the total hops, \((1 + \delta_1)\) includes the travel and decoy qubits per hop, and the additional \(+2\) corresponds to the initial Bell pair generation.

Since \(h\) Bell pairs are employed, the total qubit requirement is
\begin{equation*}
    Q_{\text{total}} = \Big[(n + 1)(1 + \delta_1) + 2\Big] \Big(\lfloor \log_2 n \rfloor  + 1\Big).
\end{equation*}

Thus, the overall qubit efficiency becomes
\begin{equation*}
    \eta_{\text{deterministic}} = \frac{1}{\Big[(n + 1)(1 + \delta_1) + 2\Big] \Big(\lfloor \log_2 n \rfloor + 1\Big)}.
\end{equation*}

For comparison, the iterative Bell-state QAV-6 protocol of~\cite{mishra2022} achieves \begin{equation*} \eta_{\mathrm{QAV\text{-}6}} = \frac{1}{\big[(n + 1)(1 + \delta_1) + 2\big] \, l}, \qquad l \le 1 + \log_2 n. \end{equation*} 

 When using  \(l = 1 + \log_2 n\), the proposed deterministic protocol has the same qubit efficiency but  reduces operational complexity by eliminating iteration. The iterative Bell-state QAV-6 protocol passes $l$ Bell pairs through the participants, but does not fully satisfy the binding property, as participants’ choices could be influenced across multiple rounds. In the proposed deterministic protocol, $h = \lfloor \log_2 n \rfloor + 1$ Bell pairs are passed \emph{once} in a single step, ensuring binding and irreversible encoding of individual veto choices.  The efficiency of Deterministic Multi-Qubit Entangled-State Protocol QAV-7 of~\cite{mishra2022}, is \begin{equation*} \eta_{\mathrm{QAV\text{-}7}} \;=\; \frac{1}{\,m + (n+1)(1+\delta_1)\,l + 1\,}, \end{equation*} where \(m\) is the size of the multi-qubit entangled resource and \(l\) is the number of qubits transfered through the participants. Although QAV-7 is deterministic and single round, the complexity of preparation of large scale entangled states and execution of participant specific unitary operations grows with the size of $n$. The proposed protocol has the benefits of Bell-state-based protocols (such as QAV-6) and single-round determinism (such as QAV-7) which can provide binding, verifiability, and correctness without involving large-scale entanglement.

\begin{table}[H]
\centering
\caption{Security Properties of QAV Protocols}
\label{tab:1}
\resizebox{\textwidth}{!}{%
\begin{tabular}{|c|c|c|c|c|c|c|c|}
\hline
\textbf{Protocol} & \textbf{Eligibility} & \textbf{Privacy} & \textbf{Binding} & \textbf{Verifiability} & \textbf{Correctness} & \textbf{Robustness} & \textbf{Commutativity} \\ \hline

QAV-6 & $\checkmark$ & $\checkmark$ & $\times$ & $\checkmark$ & $\checkmark$ & $\checkmark$ & $\checkmark$ \\

QAV-7 & $\checkmark$ & $\checkmark$ & $\checkmark$ & $\checkmark$ & $\checkmark$ & $\checkmark$ & $\times$ \\

Proposed & $\checkmark$ & $\checkmark$ & $\checkmark$ & $\checkmark$ & $\checkmark$ & $\checkmark$ & $\checkmark$ \\ \hline
\end{tabular}
}
\end{table}

\begin{table}[H]
\centering
\caption{Resource Requirements and Efficiency Comparison}
\label{tab:2}
\resizebox{\textwidth}{!}{%
\begin{tabular}{|c|c|c|c|}
\hline
\textbf{Protocol} & \textbf{Entanglement} & \textbf{Communication Rounds} & \textbf{Qubit Efficiency $\eta$} \\ \hline

QAV-6 & Bell states & Multi-round & $\big((n+1)(1+\delta_1)+2)l\big)^{-1}$ \\

QAV-7 & Multipartite & Single-round & $\big(m + (n+1)(1+\delta_1)l +1\big)^{-1}$ \\

Proposed & Bell states & Single-round & $\big((n+1)(1+\delta_1)+2)(\lfloor \log_2 n \rfloor + 1)\big)^{-1}$ \\ \hline
\end{tabular}
}
\end{table}

\section{Implementation Using Photonic Quantum Walks}  
\label{sec:implementation_using_quantum_walk}
The proposed deterministic QAV protocol can be implemented on a photonic platform by encoding two or more qubits in a single photon through its polarization and path degrees of freedom. This section outlines a feasible architecture based on current linear optical technologies and the framework of photonic quantum walks.
Quantum walks (QWs), the quantum counterparts of classical random walks, are well-suited for photonic quantum information processing due to their coherent and controllable dynamics. There are two primary variants of QWs: discrete-time (DTQWs) which describe the evolution of a
 quantum walker in discrete time steps and continuous-time (CTQWs) which describe the continuous evolution of a quantum state under a Hamiltonian that dictates its dynamical evolution~\cite{DTQW1,CTQW1,Szegedy_QW,Staggered_QW,QW_rewiew}. This work focuses on coin-based DTQWs, which are known to support universal quantum computation~\cite{Quantum_walks_review,Qc_using_QW, sangwan2025deutsch}.

A coin-based DTQW evolves in a Hilbert space \( H = H_c \otimes H_s \), where \( H_c \) denotes the coin space (polarization) and \( H_s \) represents the position space (optical path). Each step comprises a coin operator \( \hat{C}_t \), a general SU(2) rotation:
\begin{equation*} \label{Eq:coin_operator}
\hat{C}_t = e^{ip} \begin{bmatrix}
    e^{iq} \cos(\theta) & e^{ir} \sin(\theta) \\
    -e^{-ir} \sin(\theta) & e^{-iq} \cos(\theta)
\end{bmatrix}, \quad p, q, r, \theta \in \mathbb{R},
\end{equation*}
followed by a conditional shift operator:
\begin{align}
    \hat{S}_{-}^a  &= \sum_{l \in \mathbb{Z}} |a\rangle \langle a| \otimes |l-1\rangle \langle l| + \sum_{b \neq a} |b\rangle \langle b| \otimes |l\rangle \langle l|, \label{Eq:S_minus} \\
    \hat{S}_{+}^b &= \sum_{l \in \mathbb{Z}} |b\rangle \langle b| \otimes |l+1\rangle \langle l| + \sum_{a \neq b} |a\rangle \langle a| \otimes |l\rangle \langle l|. \label{Eq:S_plus}
\end{align}

The gate set \( \{\hat{I}, \hat{C}_t, \hat{S}_{\pm}^0, \hat{S}_{\pm}^1\} \) suffices to construct universal gates such as Hadamard, Phase, and CNOT~\cite{Qc_using_QW}.

In photonic DTQWs, the polarization states \( |H\rangle \) and \( |V\rangle \) encode the coin qubit, while the spatial modes \( |m\rangle \) encode the walker's position. Coin operations are implemented via a sequence of waveplates, typically a Quarter-Wave Plate (QWP), a half-wave plate (HWP), and another QWP enabling arbitrary SU(2) rotations on the polarization state. The conditional shift is realised using Beam Splitters (BSs) and Polarizing Beam Splitters (PBSs), which route photons based on their polarization. A 50:50 BS performs the transformation:
\begin{equation*}\label{Eq:BS_matrix}
BS = \frac{1}{\sqrt{2}} \begin{bmatrix} 1 & 1 \\ 1 & -1 \end{bmatrix}.
\end{equation*}

An HWP, a birefringent optical device, introduces a phase shift \( \pi \)  between orthogonal polarizations. Its Jones matrix at rotation angle \( \alpha \) is:
\begin{equation*} \label{Eq:HWP_Jones_Matrix}
\text{HWP}(\alpha) = 
\begin{pmatrix}
\cos{2\alpha} & \sin{2\alpha} \\
\sin{2\alpha} & -\cos{2\alpha}
\end{pmatrix}.
\end{equation*}

A polarization qubit encodes quantum information in the polarization state of a single photon. The basis states \( |H\rangle \) and \( |V\rangle \) represent horizontal and vertical polarizations, respectively. An arbitrary polarization qubit is given by:
\begin{equation*}
|\psi\rangle_{\text{pol}} = \alpha|H\rangle + \beta|V\rangle, \quad \text{where } |\alpha|^2 + |\beta|^2 = 1.
\end{equation*}

Similarly, the path degree of freedom can also encode a qubit, known as a path qubit. The two spatial modes \( |0\rangle \) and \( |1\rangle \) serve as the logical basis, and a general path qubit state is:
\begin{equation*}
|\psi\rangle_{\text{path}} = \alpha|0\rangle + \beta|1\rangle, \quad \text{with } |\alpha|^2 + |\beta|^2 = 1.
\end{equation*}

By combining polarization and path, two qubits can be encoded in a single photon. A general two-qubit state is:
\begin{equation*}
|\Psi\rangle_{\text{gen}} = \alpha|H\rangle|0\rangle + \beta|H\rangle|1\rangle + \gamma|V\rangle|0\rangle + \delta|V\rangle|1\rangle,
\end{equation*}
with \( |\alpha|^2 + |\beta|^2 + |\gamma|^2 + |\delta|^2 = 1 \). The logical mapping is:
\[
|H\rangle|0\rangle \equiv |00\rangle, \quad |H\rangle|1\rangle \equiv |01\rangle, \quad |V\rangle|0\rangle \equiv |10\rangle, \quad |V\rangle|1\rangle \equiv |11\rangle.
\]

To prepare the Bell state \( |\Phi^+\rangle = \frac{1}{\sqrt{2}}(|00\rangle + |11\rangle) \), the following steps are taken:
\begin{enumerate}
    \item Initialize a photon in the state \( |H\rangle|0\rangle \).
    \item Pass it through a 50:50 BS to create a path superposition:
    \[
    |\psi\rangle = \frac{1}{\sqrt{2}} \left( |H\rangle|0\rangle + |H\rangle|1\rangle \right).
    \]
    \item Insert an HWP at \( \alpha = \pi/4 \) in path \( |1\rangle \) to perform \( |H\rangle \rightarrow |V\rangle \).
    \item The resulting state is:
    \[
    |\psi\rangle_{\text{out}} = \frac{1}{\sqrt{2}} \left( |H\rangle|0\rangle + |V\rangle|1\rangle \right),
    \]
    which corresponds to \( |\Phi^+\rangle \).
\end{enumerate}

This hybrid encoding enables deterministic generation of two-qubit entanglement and gate operations with a single photon. Recent experiments~\cite{Exp_QC_using_QW} validate the feasibility of such an approach using standard linear optical components.

Each Bell pair in the protocol can thus be realised as a single photon's polarization-path state. participant operations are implemented via controlled phase shifts using Phase Shifters (PSs) which apply a phase shift of $e^{i\theta}$ to the mode in which it is implemented.

The protocol maps to the photonic architecture as follows:

\begin{enumerate}
% \item \textbf{participant Verification:}  
% Conducted via quantum digital signatures; this step is independent of the photonic setup.
\item {Initialization:}  
The Voting Authority (VA) prepares \( h \) photons, each in the entangled Bell state \( |\Phi^+\rangle \), using the BS–HWP method described earlier. Each photon encodes one Bell pair via polarization-path encoding. The VA retains one mode from each pair and sends the corresponding other mode to the first participant \( V_0 \), for each index \( a \in \{1, 2, \ldots, h\} \).

% \item \textbf{Eavesdropping Detection:}  
% Standard decoy-state techniques and polarization analysis (e.g., polarizers, single-photon detectors) are used before sending photons to participants.

\item {Sequential Vetoing:}  
Each participant receives the photon and performs one of the following actions:

\begin{itemize}
  \item \textit{Identity operation (no veto):} The photon is left unchanged.
  \item \textit{Phase gate \( P^{(a)} \) (veto):} Implemented using a PS at angle \( \pi/2^{a-1} \).
\end{itemize}

\item {Final Measurement:}  
Once all participants have done it, the photon is sent back to the VA. The interferometric recombination of the spatial modes is done by the VA, which measures the Bell state. It is achieved by placing a HWP  at \( \alpha = \pi/4 \) in the mode picked up by the participants followed by a BS, which is Bell state measurement (CNOT + Hadamard gate). Single-photon detectors are placed at the output ports of both modes to record the measurement outcomes.

For each Bell pair, if the photon is detected in the mode retained by the VA, the state is identified as \( |\Phi^+\rangle \); otherwise, detection in the opposite mode indicates a phase-flipped state \( |\Phi^-\rangle \).

\item {Veto Detection:}  
The VA analyzes the Bell measurement results to determine whether a veto occurred:
\begin{itemize}
    \item If all \( h \) outcomes are \( |\Phi^+\rangle \), then \( k = 0 \), indicating no veto was cast.
    \item If any outcome is \( |\Phi^-\rangle \), then \( k \geq 1 \), meaning at least one participant applied a veto.
\end{itemize}
This detection is performed without revealing the identity or number of vetoing participants, thereby preserving participant anonymity.

\end{enumerate}

\begin{figure}[H]
    \centering
    \includegraphics[width=0.8\linewidth]{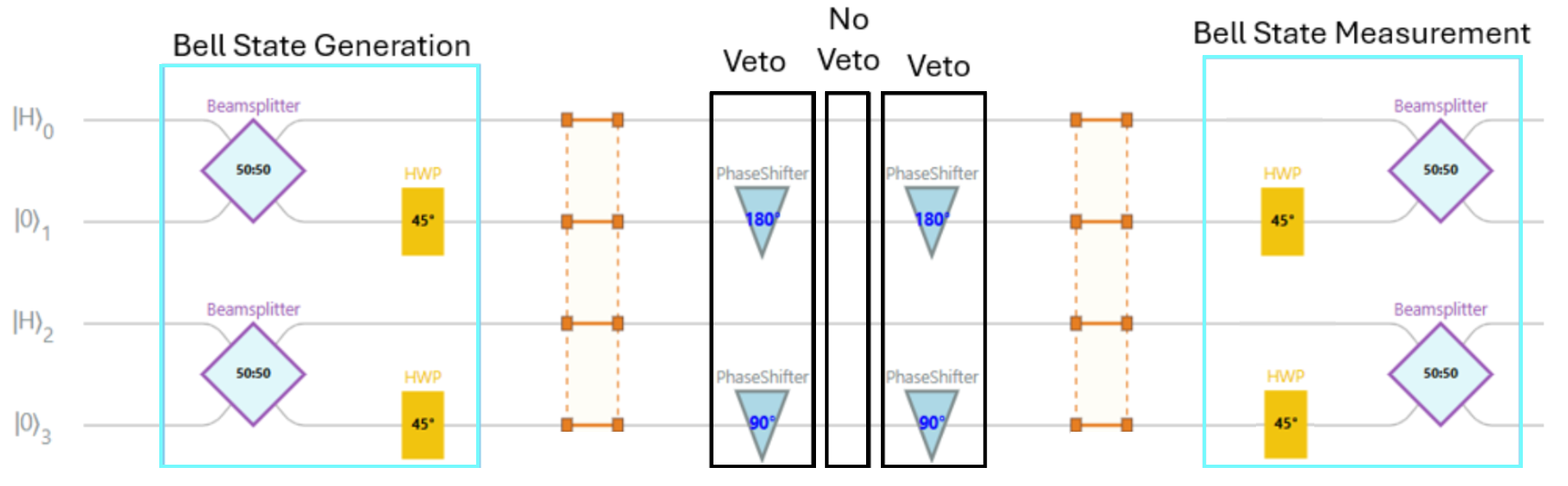}
    \caption{Photonic component based circuit demonstrating the proposed algorithm for $3$ participant where the first and the third participant exercises their veto authority.}
    \label{fig:photonic}
\end{figure}

The photonic quantum walk based implementation of the proposed algorithm for $3$ participants using photonic components is shown in the Figure \ref{fig:photonic}.

This photonic DTQW-based implementation offers a compact, low-decoherence platform for deterministic QAV. All quantum operations state preparation, vetoing, and measurement are realised using standard linear optical components, making the protocol experimentally accessible with current technologies.

\section{Conclusion}
\label{sec:conclusion}
In this work, we present a deterministic QAV protocol that facilitates conclusive veto detection in a single round of communication using only bipartite entanglement in form of Bell states. The proposed protocol requires logarithmic resources in the number of participants, does not need a multipartite entanglement and iterative communication, and thus becomes more scalable and practically implementable. As far as we know, this is the first QAV protocol to achieve deterministic single-round operation, Bell-state-based implementation, logarithmic scaling, and binding property in a single framework.

An in-depth analysis of security and efficiency shows that the protocol meets all critical requirements including anonymity, privacy, correctness, verifiability, robustness, commutativity and binding. Unlike earlier protocols, these characteristics are obtained without elevating entanglement complexity or communication overhead, resulting in a structurally simpler and operationally efficient design. In addition, we have presented a photonic implementation framework based on polarization-path encoding and discrete-time quantum walks, outlining a feasible route toward realization using current quantum optical technologies.

As such, the protocol is suitable for secure decision-making in distributed settings with strong privacy constraints, such as anonymous committee voting, privacy-preserving consensus, and quantum-secure governance. The future work may involve extending the protocol to more general voting scenarios, assessing its performance under realistic noise and decoherence models, and pursuing experimental verification of the proposed photonic scheme.

\section*{Acknowledgment}
    The quantum circuit implementations presented in this work were obtained using \textbf{Qniverse : A Unified Quantum Computing Platform} \cite{Qniverse}. This research was supported by the Ministry of Electronics and Information Technology (MeitY), Government of India, under the project titled \emph{“HPC-based Quantum Accelerators for Enabling Quantum Computing on Supercomputers”}, Grant No. 4(3)/2022-ITEA.

\onecolumn
\addcontentsline{toc}{section}{References}
\printbibliography
\end{document}